\documentclass[aps,prd,floatfix,reprint,twocolumn,reprint,amsmath,amssymb,superscriptaddress,showpacs]{revtex4-1}
\usepackage{graphicx}
\usepackage{bm}
\usepackage{amsmath}
\usepackage{dcolumn}
\usepackage{dsfont}
\usepackage{amssymb}
\usepackage{tabularx}
\usepackage{array}
\usepackage{float}
\usepackage{color}
\usepackage{epstopdf}
\usepackage{mathrsfs}
\usepackage[colorlinks, linkcolor=blue,anchorcolor=blue,citecolor=blue,urlcolor=blue]{hyperref}

\begin{document}
\title{Electrically switchable hidden spin polarization in antiferroelectric crystals }

\author{Shan Guan}
\affiliation{State Key Laboratory of Superlattices and Microstructures, Institute of Semiconductors, Chinese Academy of Sciences, Beijing 100083, China}

\author{Jun-Wei Luo}
\email{jwluo@semi.ac.cn}
\affiliation{State Key Laboratory of Superlattices and Microstructures, Institute of Semiconductors, Chinese Academy of Sciences, Beijing 100083, China}
\affiliation{Center of Materials Science and Optoelectronics Engineering, University of Chinese Academy of Sciences, Beijing 100049, China}
\affiliation{Beijing Academy of Quantum Information Sciences, Beijing 100193, China}
\begin{abstract}
Hidden spin polarization (HSP) emerges in centrosymmetric crystals where visible spin splittings in the real space can be observed because of the lack of inversion symmetry in each local sector. Starting from tight-binding models, we introduce nonsymmorphic antiferroelectric (AFE) crystals as a new class of functional materials that can exhibit strong local spin polarization. Such AFE crystals can be basically classified as in-plane and out-of-plane AFE configurations, and can be reversibly switched to the ferroelectric phase by an electric field to manifest global spin splittings, enabling a nonvolatile electrical control of spin-dependent properties. Based on first-principle calculations, we predict the realization of strong HSP in the AFE phase of a newly-discovered two-dimensional materials, quintuple-layer (QL) LiBiO$_2$. Furthermore, the spontaneous electric polarization ($\sim$ 0.3 nC/m) and the transition barrier as well as the tunable spin polarization of QL-LiBiO$_2$ are discussed.
\end{abstract}

\maketitle
Spin-orbit coupling (SOC) causes a momentum-dependent splitting of electron states in noncentrosymmetric nonmagnetic solids and gives rise to two conventional types of spin texture: the Dresselhaus-type due to bulk inversion asymmetry~\cite{Dresselhaus} and the Rashba-type~\cite{Rashba1960,Rashba1984} in two-dimensional (2D) heterostructures due to structural inversion asymmetry. Such SOC was recently discovered to be hidden in certain centrosymmetric crystals~\cite{Zhang2014,Partoens2014} in which noncentrosymmetric sectors (or sublattices) connected by inversion experience a noncentrosymmetric environment and have a local spin-polarization that is hidden globally by its inversion-partner with an opposite polarization. According to the site symmetry of the noncentrosymmetric sublattices, we can classify the hidden spin-polarization (HSP) into Rashba type (R-2) or Dresselhaus type (D-2) depending on whether the site inversion asymmetry or the site dipole field plays a crucial role in forming the hidden spin texture. This discovery has triggered much attention on broader hidden physical effects, such as spin-orbit torque~\cite{Zielezny2014}, circular polarization~\cite{Liu2015}, orbital polarization~\cite{Ryoo2017}, Berry curvature~\cite{Cho2018}, unconventional superconductor~\cite{Liu2017,Gotlieb2018}, and so on. 

Most centrosymmetric crystals are constituted by inversion-paired noncentrosymmetric sectors and thus would expect to show HSP. However, the strength of the local spin-polarization of a given sector could be seriously suppressed by the interaction between sectors due to the degeneracy of their electron states arising, respectively, from inversion-paired sectors ~\cite{Zhang2014,Li2018}. Fortunately, in previous work~\cite{Yuan2019} we have demonstrated that there exist certain symmetry operations, such as nonsymmorphic symmetry operation, could prevent electron states from such inter-sector coupling to realize strong HSP in centrosymmetric crystals containing those symmetry operations. Therefore, the quest of centrosymmetric materials for strong HSP is an interesting task. In addition, for exploration of HSP in functional device applications, developing routes towards the control of local spin-polarization is highly desirable, especially by means of external electric fields owing to its compatibility with current semiconductor technology. 

A promising direction to tackle this challenging task is to utilize antiferroelectric (AFE) materials~\cite{Ishizaka2011,DiSante2013,Tagantsev2013,Hao2014,Monserrat2017}. Antiferroelectricity is closely related to ferroelectricity; the relation between antiferroelectricity and ferroelectricity is analogous to the relation between antiferromagnetism and ferromagnetism. Ferroelectricity results from relative shifts of negative and positive ions that induce an array of electric dipoles all point in the same direction and generate a net macroscopic spontaneous polarization, as a result of phase transition from the paraelectric (PE) phase to the ferroelectric (FE) phase. This can be contrasted with an antiferroelectric, in which adjacent dipoles oriented in opposite (antiparallel) directions (the dipoles of each orientation form interpenetrating sublattices) and thus the total macroscopic spontaneous polarization is zero since the adjacent dipoles cancel each other out. Given that the local electric dipoles with opposite orientations are connected by inversion in centrosymmetric AFE materials~\cite{Kittel1951,Shirane1951,Kim2006,Kan2010}, one naturally wonders whether such local electric dipoles could serve as site inversion-asymmetric potentials distributed in individual sectors to render a strong HSP. Particularly, AFE lattices with specific nonsymmorphic symmetry are preferred, owing to their advantage in protection of a strong HSP by suppressing the interaction between different sectors~\cite{Yuan2019,Zhang2019}.

\begin{figure}[t!]
\centerline{\includegraphics[width=0.49\textwidth]{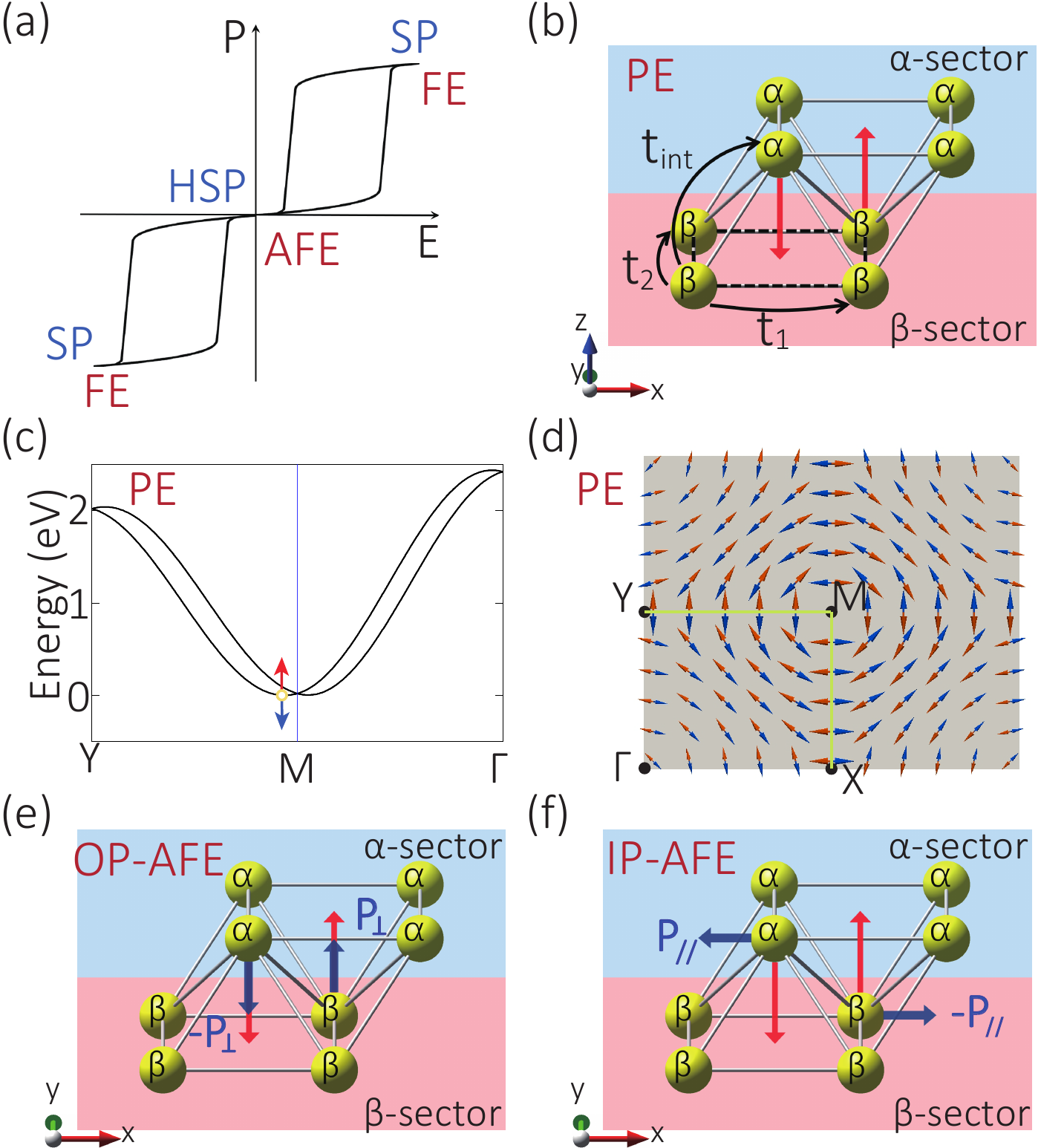}}
\caption{(a) Schematic representation of tunable electric polarization ($P$) by applying an external electric field ($E$) in an AFE crystal. (b) Side view of the 2D nonsymmorphic rectangle lattice (PE phase). The shaded regions in blue and red indicate a pair of sectors labeled $\alpha$ and $\beta$, respectively. The dashed line indicates the unit cell. $t_1$ ($t_{2}$) is the nearest neighbor hopping parameter along $x$ ($y$) within the same sector, while $t_{int}$ is the nearest neighbor hopping between different sectors. The red arrows denote the intrinsic local dipole fields on each atom. (c) Band structure for PE phase. (d) Projected spin texture onto sector $\alpha$ (blue) and $\beta$ (red) for PE phase. Plots of polarization configuration for (e) OP-AFE and (f) IP-AFE, where the blue arrows represent the $P$ directions. See Ref.~\cite{sp} for used parameters in calculation.}
\label{fig1}
\end{figure}

In this Rapid Communication, we show that nonsymmorphic AFE crystals can indeed achieve both the objectives mentioned above. Specifically, the AFE systems, which possess an antipolar state and two opposite polar states~\cite{Kittel1951,Monserrat2017}, offer an easy tuning of their local spin-polarization via an external electric field as enforcing a transition from an AFE phase (with HSP) to a FE phase (with visible spin-polarizations), as shown schematically in Fig.~\ref{fig1}(a). In the following, AFE-HSP denotes the AFE materials in which the antipolar state exhibits strong HSP. For layered AFE-HSP, we further find two different configurations, regarding the orientation of the local electric dipoles with respect to the layer plane: one is in-plane (IP) direction and the other is out-of-plane (OP) direction. One may expect distinct hidden spin patterns for these two configurations.

We use simple 2D lattice models to illustrate the effect of switchable electric polarization on spin texture by taking an example of nonmagnetic system for it has the time-reversal symmetry $\mathcal{T}$. We first examine the PE phase in the absence of electric dipoles arising from relative shifts of negative and positive ions that occurs in the AFE phase. However, in the PE phase, intrinsic dipole fields emerge unavoidably, which will also lead to HSP. Fig.~\ref{fig1}(b) shows the side view of a nonsymmorphic 2D lattice, where two identical atoms $\alpha$ and $\beta$ in $\alpha$- and $\beta$-sectors connected by inversion are displaced inward- and outward-facing (buckling), respectively, producing intrinsic local dipole fields (with opposite directions) perpendicular to the layered plane, as illustrated by red arrows in Fig.~\ref{fig1}(b). Such lattice has a p4/nmm layer group holding two screw axis operations $\{\mathcal{C}_x|\frac{1}{2}0\}$ and $\{\mathcal{C}_y|0\frac{1}{2}\}$. Taking $\{|\alpha \rangle,|\beta \rangle \}\otimes\{\uparrow ,\downarrow\}$ as the basis with an assumption of single $s$-orbital in each atom contributing to the low-energy spectrum (four bands including spin), the Hamiltonian in $\bm k$ space can be written as~\cite{sp,Liu2011,Zhang2019}
\begin{equation}\label{PESOC}
\begin{split}
\mathcal{H}^{\text{PE}}=&(t_1cosk_x+t_2cosk_y)+t_{int}cos\frac{k_x}{2}cos\frac{k_y}{2}\tau_x  \\
                        &\otimes \sigma_0 +\lambda \tau_z \otimes(sink_x \sigma_y-sink_y  \sigma_x), 
\end{split}
\end{equation}
where the Pauli matrices $\tau$ and $\sigma$ are for the sector and the spin degrees of freedom, respectively. The first term $\mathcal{H}^{\text{PE}}_0$ describes the hopping within each sector with hopping parameter $t_1$ ($t_{2}$) for the nearest neighbor interactions along $x$ ($y$), and the second term $\mathcal{H}^{\text{PE}}_{int}$ for the interaction between inversion-paired sectors with $t_{int}$ for the nearest neighbor hopping. Whereas, the third term $\mathcal{H}^{\text{PE}}_{so}$ represents the SOC interaction resulting from the intrinsic local dipole fields due to the out-of-plane buckling of $\alpha$ and $\beta$ atoms~\cite{sp}. It is straightforward to observe that the inter-sector interaction vanishes $\mathcal{H}^{\text{PE}}_{int}=0$ at the BZ boundary ($k_x$=$\pi$ or $k_y$=$\pi$), leading to commutativity of the Hamiltonian $\mathcal{H}^{\text{PE}}$ with the sector measurement operator $\tau_{z}\otimes\sigma_{0}$. Such commutativity yields wavefunction segregated on a single sector instead of spreading over both inversion-paired sectors, which guarantees strong HSP in centrosymmetric crystals. $\mathcal{H}^{\text{PE}}_{so}$ also suggests that the HSP at the M point is R-2 but at the X and Y points are D-2. Indeed, our calculation shown in Fig.~\ref{fig1}(c-d) indicates that the spin-polarizations from the $\alpha$- and $\beta$-sector have opposite directions, and exhibit a helical spin texture around M point but a non-helical one around X and Y points. Therefore, a strong R-2 accompanied by D-2 is obtained for the PE phase, consistent with previous findings~\cite{Yuan2019,Zhang2019}.

\begin{figure}[!b]
\centerline{\includegraphics[width=0.5\textwidth]{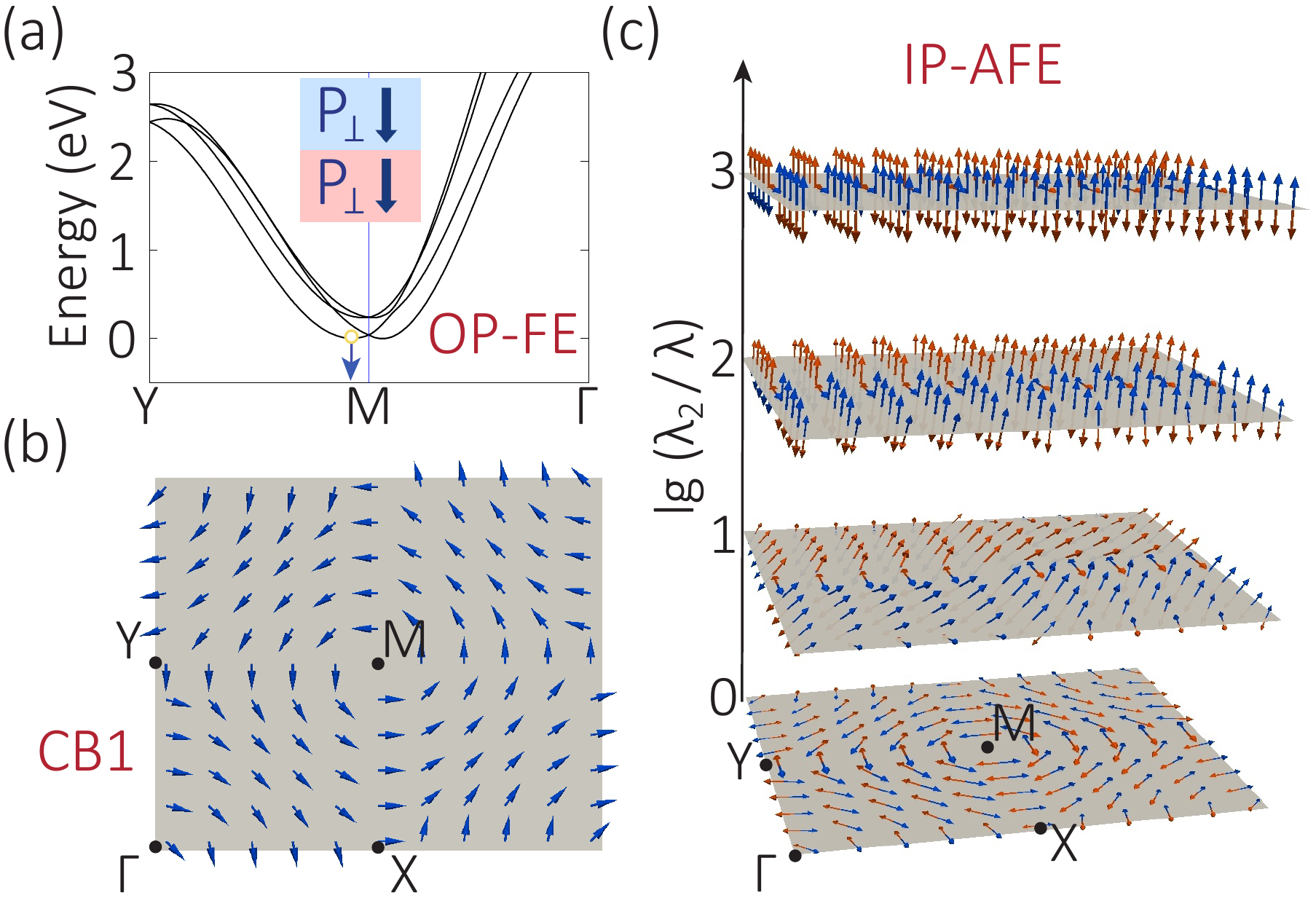}}
\caption{(a) Band structure for OP-FE. Inset is the corresponding plot of polarization configuration. (b) The spin pattern of CB1 for OP-FE. (c) Projected spin-polarizations as a function of $lg(\lambda_2/\lambda)$ for IP-AFE.}
\label{fig2}
\end{figure}

In actual materials, except $\alpha$ and $\beta$ atoms, each sector can contain other negative and positive ions whose relative shifts generate an additional local dipole field felt oppositely by $\alpha$ and $\beta$ atoms and cause a phase transition from PE to AFE. We then turn to the AFE phase containing such distortion-induced additional local electric dipoles. When distortion-induced local dipoles $P_{\perp}$ pointing out the plane as shown in Fig.~\ref{fig1}(e), we obtain the OP-AFE configuration, in which all the symmetry operations of the PE phase are preserved. As a result, relative to PE Hamiltonian, only the effective SOC term need to be altered by $\mathcal{H}^{\text{OP-AFE}}_{so}$=$(\lambda+\lambda_{1})\tau_z\otimes(sink_x\sigma_y-sink_y\sigma_x)$, where $\lambda_{1}$ refers to the coupling strength stemming from the distortion-induced dipoles and hence is proportional to $P_{\perp}$. $\mathcal{H}^{\text{OP-AFE}}_{so}$ is so similar to $\mathcal{H}^{\text{PE}}_{so}$ that OP-AFE preserves strong HSP as in the PE phase. Furthermore, upon application of an external electric field antiparallel to $P_{\perp}$, we can reverse the direction of the distortion-induced electric dipole in the $\beta$-sector from $z$ to $-z$ and thus cause an AFE-to-FE phase transition along with the breaking of the global inversion symmetry. Thus, an on-site energy $E_{0}\tau_z \otimes \sigma_0$ should be added to the Hamiltonian and the corresponding SOC term for the OP-FE now reads $\mathcal{H}^{\text{OP-FE}}_{so}$=$(\lambda\tau_z+\lambda_{1} \tau_0)\otimes(sink_x\sigma_y-sink_y\sigma_x)$. Fig.~\ref{fig2}(a-b) show the band structure and corresponding spin-polarization for OP-FE. We see that the two-fold degeneracy of each band is now lifted and a conventional Rashba spin texture is emergent, manifesting a reversible and nonvolatile manipulation of local spin-polarizations.

When distortion-induced local dipoles $P_{\parallel}$ parallel the layer plane, say along the $\pm x$-direction [see Fig.~\ref{fig1}(f)], we obtain the IP-AFE configuration. The symmetry of the lattice is lowered and preserves only one nonsymmorphic operation, i.e., $\{\mathcal{C}_y|0\frac{1}{2}\}$. The hopping between sectors becomes anisotropic, which yields the inter-sector interaction Hamiltonian $\mathcal{H}^{\text{IP-AFE}}_{int}$=$cos\frac{k_y}{2}[(t_{int1}e^{i\frac{k_x}{2}}+t_{int2}e^{-i\frac{k_x}{2}})(\tau_x+i\tau_y)\otimes \sigma_0$+h.c.]~\cite{sp}. Besides the intrinsic out-of-plane local dipoles, the presence of $P_{\parallel}$ results in an additional SOC term $\mathcal{H}^{\text{IP-AFE}}_{so}$=$-\lambda_2 sink_y \tau_z \otimes \sigma_z$. Therefore, the total Hamiltonian for IP-AFE becomes
\begin{equation}\label{IPSOC}
\mathcal{H}^{\text{IP-AFE}}=\mathcal{H}^{\text{PE}}_0+\mathcal{H}^{\text{IP-AFE}}_{int}+\mathcal{H}^{\text{PE}}_{so}+\mathcal{H}^{\text{IP-APE}}_{so}.
\end{equation}
Due to the vanishing of the inter-sector interaction as a result of the preserved $\{\mathcal{C}_y|0\frac{1}{2}\}$ combined with time-reversal symmetry $\mathcal{T}$~\cite{Yuan2019,Tao2017}, IP-AFE configuration possesses a strong HSP along the Y-M $k$-line. Fig.~\ref{fig2}(c) exhibits the sector-decomposed spin-polarization as a function of $lg(\lambda_2/\lambda)$, which also reflects the ratio between $P_{\parallel}$ and the intrinsic local dipole. In the limit of $P_{\parallel}=0$, IP-AFE configuration is equal to a PE phase with in-plane spin orientations as discussed above. As increasing $P_{\parallel}$, the sector-dependent spin has a tendency to orient towards the $z$- or $-z$-direction. Interestingly, it finally reaches a Zeeman-like hidden spin pattern, an unrecognized effect, when the distortion-induced $P_{\parallel}$ is so large that intrinsic local dipoles are negligible. Based on a simple tight-binding model, we have uncovered all interesting spin-polarization properties for AFE crystals.

\begin{figure}[!t]
\centerline{\includegraphics[width=0.5\textwidth]{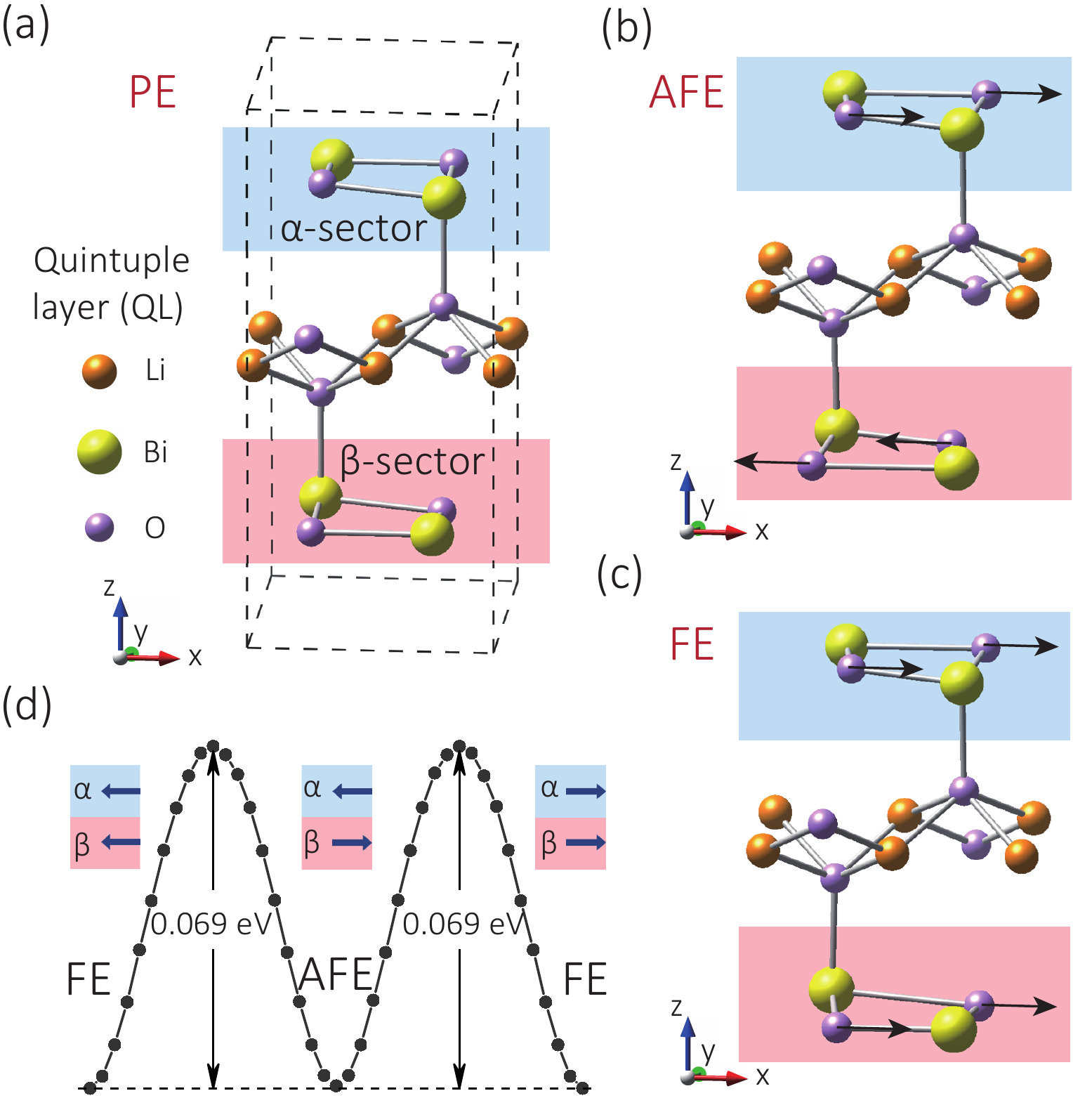}}
\caption{(a) Crystal structure of PE QL-LiBiO$_2$. The two shaded real-space sectors indicate the inversion partners that are used for spin projection. Side views of the (b) AFE and (c) FE phases of QL-LiBiO$_2$. AFE Phase can be considered as distorted from PE with opposite distortions for O atoms in different BiO layers, while FE Phase is the case with identical distortions. (d) Energy profile of the polarization reversal of FE phase, during which an intermediate state, AFE phase, arises.}
\label{fig3}
\end{figure}

We now turn to examine the AFE-HSP in an actual material---the 2D quintuple (QL) LiBiO$_2$ in which the Bi atoms share the same lattice as our model. Bulk phase of LiBiO$_2$ crystallizes in a layered orthorhombic structure with space group No.72 (Ibam)~\cite{Greaves1989}. Fig.~\ref{fig3}(a) shows the crystal structure of quintuple monolayer from a bulk LiBiO$_2$, which is composed of five atomic layers stacking in the sequence of BiO-O-Li-O-BiO. Without lattice distortion, this QL-LiBiO$_2$ is in a PE phase. From the phonon spectrum of QL-LiBiO$_2$, we observe two pronounced soft optical modes at the $\Gamma$ point~\cite{sp}, responsible for out-of-phase movements of O atoms in two BiO layers against each other along the $x$-direction~\cite{sp}. One mode is for the distortions towards a centrosymmetric structure (i.e., AFE phase) and the other is towards a noncentrosymmetric structure (i.e., FE phase) as shown in Fig.~\ref{fig3}(b-c). We validate that both AFE and FE phases are dynamically stable since no imaginary frequency is presented in their calculated phonon dispersions~\cite{sp}. Because the relative shifts of O atoms occur within the plane of BiO layer, the distortion-induced local dipoles in both AFE and FE phases are in IP configurations. According to the calculated total energy listed in Table~\ref{tab:tab1}, we find that the QL-LiBiO$_2$ is stabilized in the FE phase with a total energy slightly lower than that of AFE phase ($\sim$0.4 meV/u.c.). Therefore, the AFE and FE phases are nearly degenerate.

\begin{table}[!t]
\caption{The basic properties of QL-LiBiO$_{2}$ are listed. $a$ and $b$ are the lattice constants, $E_{total}$ is the total energy with respect to the FE phase, and $E_{gap}$ are the indirect band gaps.}
\label{tab:tab1}
\begin{tabular}{p{1.8cm}<{\centering}|p{1cm}<{\centering}| p{1cm}<{\centering}| p{2cm}<{\centering}| p{1.8cm}<{\centering}}
  \hline\hline
   phase &  $a$({\AA}) & $b$({\AA}) & $E_{total}$(meV)  & $E_{gap}$(eV)  \\ \hline
    PE & 4.736 & 5.349 &  27.45  & 2.167 \\ \hline
   AFE & 4.901 & 5.279 &  0.39  & 2.324 \\ \hline
    FE & 4.901 & 5.279 &  0  & 2.318 \\
  \hline\hline
\end{tabular}
\end{table} 

\begin{figure}[!b]
\centerline{\includegraphics[width=0.49\textwidth]{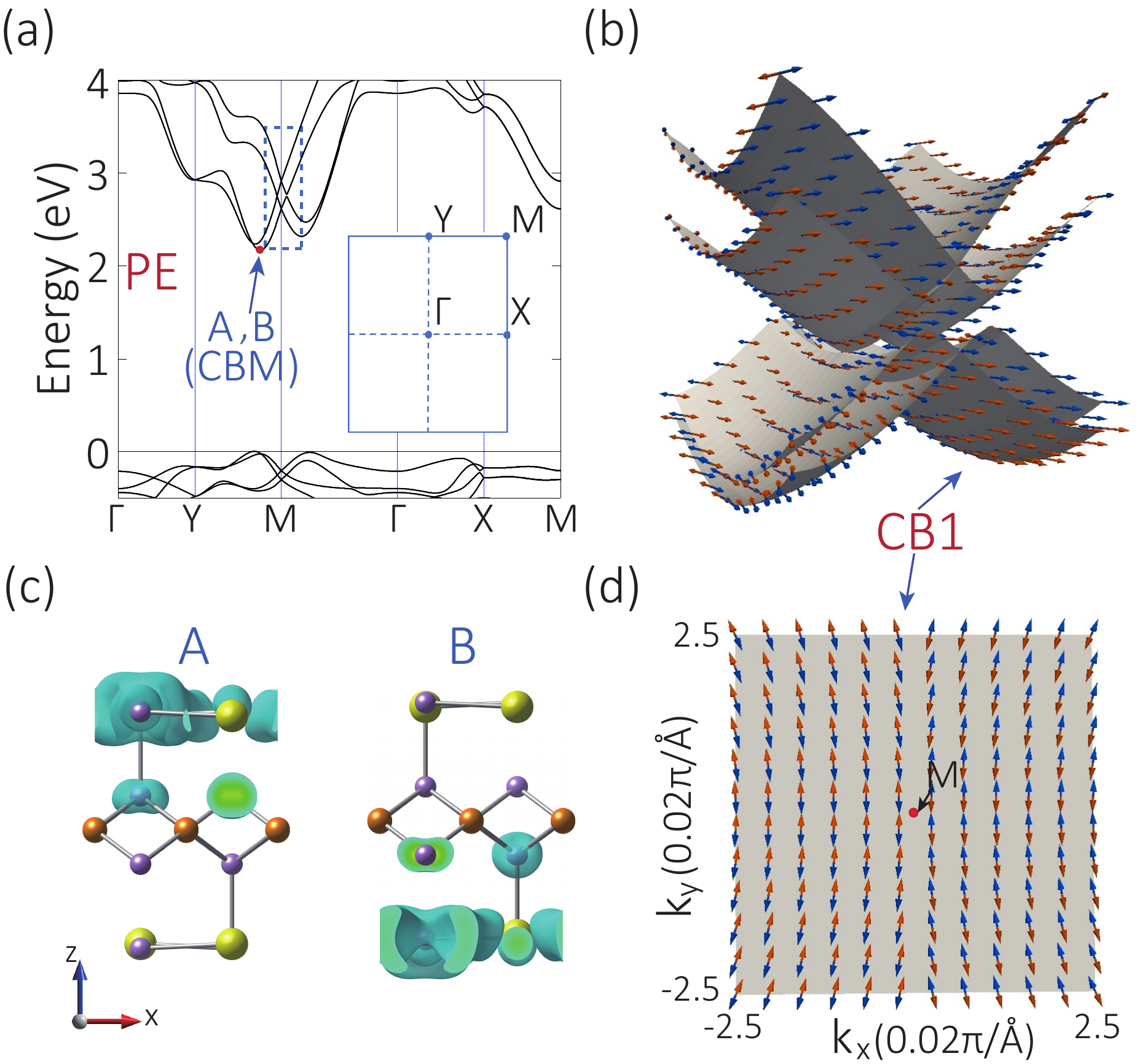}}
\caption{(a) Band structure of PE QL-LiBiO$_2$ with SOC. Inset is the first BZ. (b) Spin-polarizations are illustrated for the lowest eight conduction band states indicated by the dashed box in (a). The blue and red arrows represent the projection on the $\alpha$- and $\beta$-sectors, respectively. (c) Charge density distribution plotted for states A and B. (d) Corresponding 2D plot of spin-polarizations for CB1.}
\label{fig4}
\end{figure}

By using the Berry phase method~\cite{King-Smith1993,Resta1994}, we indeed find the ferroelectricity in QL-LiBiO$_2$ with an IP polarization of 0.301 nC/m, comparable to that of ultra-thin SnTe ($\sim$0.194 nC/m)~\cite{Chang2016,Liu2018d} and In$_2$Se$_3$ ($\sim$0.259 nC/m)~\cite{Ding2017,Zhou2017}. We obtain that the kinetics pathway of switching polarization undergoes an intermediate state (,i.e., the AFE phase), as shown in Fig.~\ref{fig3}(d). Therefore, the switching of polarization is achievable through the reverse of the electric polarization in each sector in tandem. The energy barrier is approximately 69 meV, which is much larger than that for SnTe ($\sim$9.0 meV)~\cite{Wan2017} and is close to the value for In$_2$Se$_3$ ($\sim$66 meV)~\cite{Ding2017}, implying robust ferroelectricity in QL-LiBiO$_2$. Moreover, \textit{ab-initio} molecular dynamics simulation shows that the spontaneous polarizations are sustained above 300 K~\cite{sp}, indicating QL-LiBiO$_2$ having room-temperature ferroelectricity.

\begin{figure}[!t]
\centerline{\includegraphics[width=0.49\textwidth]{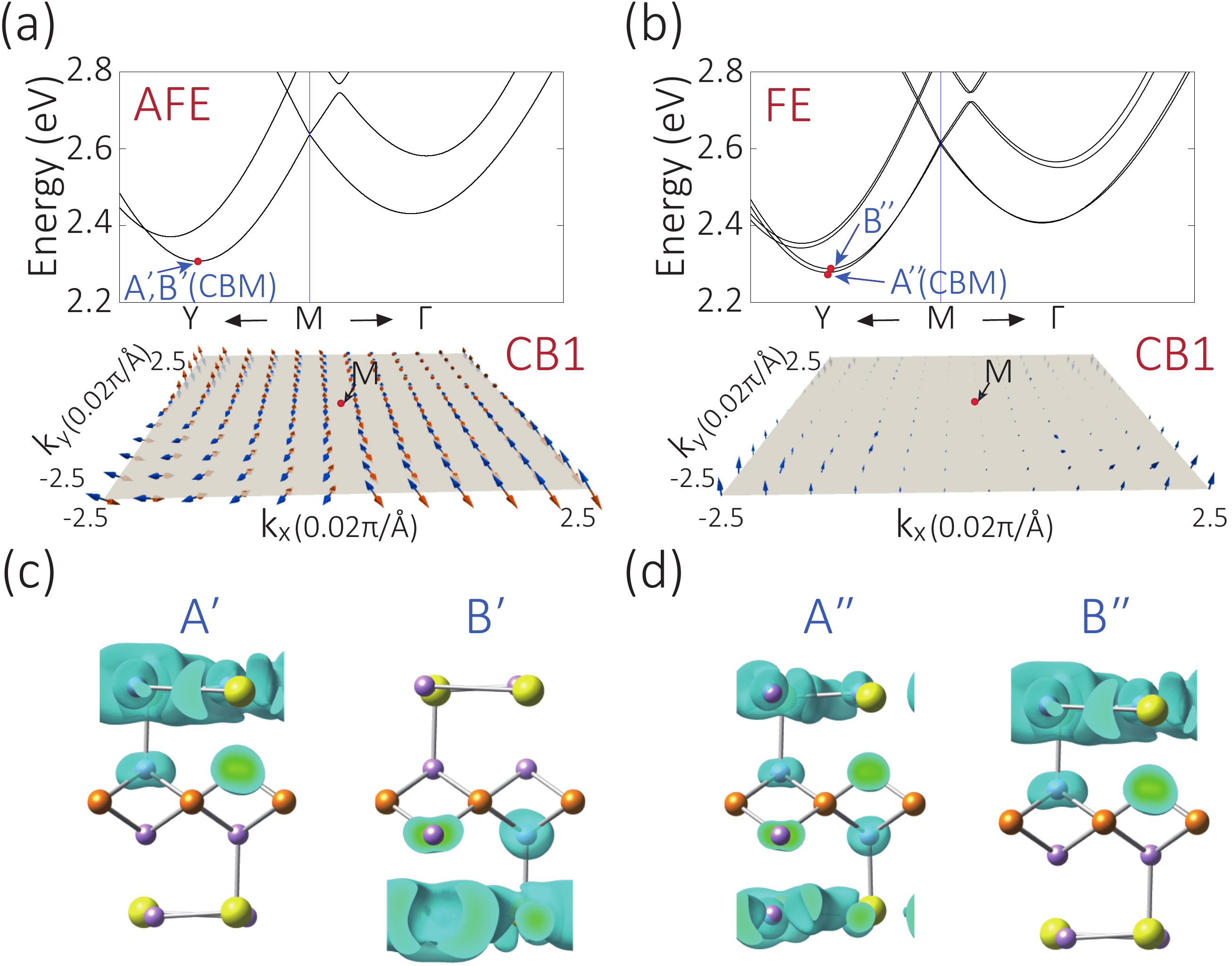}}
\caption{Band structures of QL-LiBiO$_2$ with SOC for (a) AFE and (b) FE phases, respectively. The bottom panel is the corresponding local spin-polarizations near the M point for CB1. Charge density distribution plotted for the states (c) A$^{'}$ and B$^{'}$ of AFE phase and (d) A$^{''}$ and B$^{''}$ of FE phase.}
\label{fig5}
\end{figure}

To inspect the local spin-polarization in QL-LiBiO$_2$, we investigate the PE phase where the distortion-induced dipoles vanish but still owning the intrinsic local dipoles. These intrinsic local dipoles are perpendicular to the layered plane. The PE phase has nonsymmorphic structure with a space group of Pbcm, including an inversion center and a screw axis operation $\widetilde{\mathcal{C}}_y=\{C_y|\frac{1}{2}\frac{1}{2}\}$. These two symmetry operations are critical for strong HSP. Fig.~\ref{fig4}(a) shows the band structure of the PE phase, exhibiting an indirect band gap around 2.17 eV, which is in good agreement with the experimental result (2.04 eV) of ultra-thin LiBiO$_2$ nanosheets~\cite{Zhou2019}. We find that all energy bands are doubly degenerate as expected for centrosymmetric materials. The lowest eight conduction bands (CB) marked by the dashed box are mainly from the two outermost BiO layers, which experience oppositely intrinsic local dipoles. Fig.~\ref{fig4}(b) exhibits the calculated sector-decomposed local spin-polarization in the vicinity of the M point. It clearly shows that the degenerate band states projected onto $\alpha$- or $\beta$-sectors have opposite spin polarizations, demonstrating the existence of HSP. Furthermore, the 2D spin pattern of CB1 in Fig.~\ref{fig4}(d) displays a D-2 spin texture. Particularly, due to the joint operation $\widetilde{\mathcal{C}}_y\mathcal{T}$, the relevant spin wavefunctions along the Y-M path should be segregated spatially on just one of the two separate sectors~\cite{Yuan2019,Tao2017}. By analyzing the spatial distribution of the CBM states A and B [see Fig.~\ref{fig4}(c)], we indeed confirm that the state A (B) has wavefunction confined in sector $\alpha$ ($\beta$), demonstrating the characteristic feature of strong HSP.

Next, we study the distorted structures of QL-LiBiO$_2$ where the paired IP spontaneous polarizations offer additional horizontal dipole fields for the BiO atomic layers in different sectors. For the AFE phase, the electric polarizations are opposite, and hence the inversion and screw axis operation $\widetilde{\mathcal{C}}_y$ are maintained. Consequently, the bands in Fig.~\ref{fig5}(a) are still two-fold degenerate, and the states on Y-M have their wavefunctions also segregated on a definite sector [see Fig.~\ref{fig5}(c)]. Compared with the PE phase, the difference lies in that the $z$-component of spin vector induced by the additional dipole fields becomes dominant over its IP-component when it moves away from the M point. However, for the FE phase where the paired electric polarizations share the same orientations, the band degeneracy is lifted because of the inversion symmetry breaking. Then we have two splitted states A$^{''}$ and B$^{''}$ [see Fig.~\ref{fig5}(b)]. And the global spin-polarizations without any compensation for CB1 are obtained. Furthermore, it is shown in Fig.~\ref{fig5}(d) that the resulting wavefunction distribution of states A$^{''}$(B$^{''}$) is essentially delocalized over both sectors.  

In conclusion, we have theoretically investigated the HSP in AFE systems. We present nonsymmorphic tight-binding models for realizing strong HSP in different AFE configurations, where the local spin-polarizations are tunable via polarization switching. Interestingly, we discover a novel Zeeman-like HSP. Moreover, we identify QL-LiBiO$_2$ as a realistic example to host strong AFE-HSP. We also demonstrate the intrinsic ferroelectricity and antiferroelectricity in QL-LiBiO$_2$. Our findings thus open new perspectives for the HSP research and offer a concrete material platform to explore the intriguing physics of AFE-HSP.

\begin{acknowledgments}
The work is supported by the National Key R\&D Program of China (Grant No. 2018YFB2202802), the NSF of China (Grants No.11904359), and the Strategic Priority Research Program of Chinese Academy of Sciences (Grant No. XDB30000000). 
\end{acknowledgments}

%

\end{document}